\documentclass[aps,superscriptaddress,reprint,
               colorlinks=true,
               linkcolor=black,
               citecolor=black,
               urlcolor=black]{revtex4-2}

\usepackage{graphicx}
\usepackage{dcolumn}
\usepackage{bm}
\usepackage{amsmath,amssymb}
\usepackage{xcolor}
\usepackage{pifont}
\usepackage[bottom]{footmisc}
\usepackage{booktabs}
\usepackage{multirow}
\usepackage{array}
\usepackage{caption}
\DeclareCaptionLabelSeparator{pipe}{\,\textbar\,}
\usepackage{float}
\usepackage[rightcaption]{sidecap}
\sidecaptionvpos{figure}{t}
\usepackage{titlesec}
\titleformat{\section}
  {\normalfont\bfseries\normalsize}
  {}{0em}{}
\titlespacing*{\section}
  {0pt}{10pt plus 2pt minus 2pt}{2pt}
\titleformat{\subsection}
  {\normalfont\bfseries\normalsize}
  {}{0em}{}
\titlespacing*{\subsection}
  {0pt}{4pt plus 1pt minus 1pt}{1pt}
\usepackage{orcidlink}

\begin{document}
\preprint{APS/123-QED}

\title{From Narrow-Gap Semiconductor to Metallic Altermagnet: \\ Optical Fingerprints of Co-Doped FeSb$_2$}

\author{R.~Mathew Roy\orcidlink{0009-0003-2842-9044}}
   \thanks{renjith.mathew-roy@pi1.uni-stuttgart.de}
\affiliation{1.~Physikalisches Institut, Universit\"at Stuttgart, Pfaffenwaldring 57,
70569 Stuttgart, Germany}

\author{M.~Povolotskiy\orcidlink{0009-0003-2963-8294}}
\affiliation{1.~Physikalisches Institut, Universit\"at Stuttgart, Pfaffenwaldring 57,
70569 Stuttgart, Germany}

\author{J.~Kirschke}
\affiliation{Experimental Physics IV, Department of Physics and Astronomy,
Ruhr University Bochum,
44801 Bochum, Germany}

\author{C.~Prange\orcidlink{0009-0001-6126-9498}}
\affiliation{1.~Physikalisches Institut, Universit\"at Stuttgart, Pfaffenwaldring 57,
70569 Stuttgart, Germany}

\author{Y.~Xia}
\affiliation{Institute of Theoretical Physics, Chinese Academy of Sciences, Beijing 100190, China}

\author{V.~Sundaramurthy\orcidlink{0009-0001-0112-0720}}
\affiliation{Max Planck Institute for Solid State Research, Heisenbergstraße 1, 70569 Stuttgart, Germany}

\author{P.~Puphal\orcidlink{0000-0003-3574-2170}}
\affiliation{Max Planck Institute for Solid State Research, Heisenbergstraße 1, 70569 Stuttgart, Germany}

\author{M.~Pinteri\'{c}}
\affiliation{1.~Physikalisches Institut, Universit\"at Stuttgart, Pfaffenwaldring 57,
70569 Stuttgart, Germany}
\affiliation{Faculty of Civil Engineering, Transportation Engineering and Architecture, University of Maribor, SI-2000 Maribor, Slovenia}

\author{M.~van~de~Loo}
\affiliation{Experimental Physics IV, Department of Physics and Astronomy,
Ruhr University Bochum,
44801 Bochum, Germany}

\author{A.~Kreyssig\orcidlink{0009-0003-8474-2879}}
\affiliation{Experimental Physics IV, Department of Physics and Astronomy,
Ruhr University Bochum,
44801 Bochum, Germany}

\author{T.~Zhang}
\affiliation{Institute of Theoretical Physics, Chinese Academy of Sciences, Beijing 100190, China}

\author{A.~E.~Böhmer\orcidlink{0000-0001-6836-2954}}
\affiliation{Experimental Physics IV, Department of Physics and Astronomy,
Ruhr University Bochum,
44801 Bochum, Germany}

\author{M.~Dressel\orcidlink{0000-0003-1907-052X}}
\affiliation{1.~Physikalisches Institut, Universit\"at Stuttgart, Pfaffenwaldring 57,
70569 Stuttgart, Germany}

\author{M.~Wenzel\orcidlink{0009-0000-8813-2080}}
  \thanks{maxim.wenzel@pi1.uni-stuttgart.de}
\affiliation{1.~Physikalisches Institut, Universit\"at Stuttgart, Pfaffenwaldring 57,
70569 Stuttgart, Germany}

\begin{abstract}
The realization of bulk metallic altermagnetism has remained elusive despite the growing number of candidate materials. Here, we present evidence that moderate cobalt substitution ($\sim 15$\%) drives the correlated narrow-gap semiconductor FeSb$_2$ into a metallic altermagnetic state persisting up to room temperature. The infrared optical conductivity reveals low-energy interband transitions near 0.1~eV that emerge upon doping and grow with Co concentration. Density functional theory calculations show that these transitions originate exclusively from altermagnetic spin ordering, with spin split bands ($\sim$0.2~eV) of non-relativistic origin, together with spin-orbit coupling induced band splitting of the order of $\sim$5~meV near the Fermi level. Co substitution further leads to Fano lineshapes and mode mixing in the infrared-active phonons, reflecting enhanced electron–phonon coupling and local inversion symmetry breaking, while leaving the altermagnetic spin symmetry intact. Our results establish carrier-tuned FeSb$_2$ as a platform for exploring metallic $d$-wave altermagnetism and its coupling to lattice degrees of freedom.

\end{abstract}

\date{\today}%
\maketitle
%-------------------------------------------------------------------------------------------------------------------------------------------------------

\section*{Introduction}
Altermagnets constitute a class of collinear antiferromagnets in which electronic bands exhibit momentum-dependent ``alternating'' spin splitting protected by crystal rotational symmetries, yielding a ferromagnetic-like spin splitting while maintaining a vanishing net magnetization \cite{vsmejkal2022emerging,vsmejkal2022beyond}. As a result, altermagnets can generate strongly spin-polarized currents without any macroscopic magnetization, paving new avenues for spintronic devices beyond conventional ferro-/antiferromagnets \cite{jungwirth2016antiferromagnetic,guo2025spin,bai2024altermagnetism}.

While many well-known materials satisfy the symmetry criteria for altermagnetism~\cite{wei2024crystal,vsmejkal2022beyond}, the direct identification of altermagnetic order has only recently been achieved. To date, the most direct hallmark is the lifting of Kramers degeneracy in the absence of strong relativistic effects, observed unambiguously by spin- and angle-resolved photoemission spectroscopy (ARPES) in MnTe~\cite{osumi2024observation} and CrSb~\cite{ding2024large,yang2025three}, as well as thin films of CrSb~\cite{santhosh2025altermagnetic} and RuO$_2$~\cite{fedchenko2024observation}. While surface-sensitive probes inferred metallic $d$-wave altermagnetism in KV$_2$Se$_2$O, neutron scattering and density functional theory calculation clearly evidences conventional antiferromagnetic order \cite{jiang2025metallic,sun2025neutronscattering-KV2Se2O, thapa2026altermagnetism}. Hence, the realization of a metallic bulk $d$-wave 
altermagnet remains to be established.

From an experimental perspective, altermagnetic band symmetry can induce unusual optical effects that directly reflect the underlying electronic structure. In particular, non-relativistic spin splittings give rise to circular dichroism and Kerr rotation under specific symmetry conditions \cite{liebman2026strain,sivianes2025optical,jiaxin2026symmetry,chen2025magneto}, while interband optical transitions provide a highly sensitive probe for distinguishing between magnetic and nonmagnetic band structures \cite{wenzel2025fermi}. These optical methods offer bulk sensitivity and can probe a single domain even in a multidomain sample, hence avoiding extrinsic artifacts common in transport measurements \cite{chu2025third,galindez2025revealing,gonzalez2024anisotropic} or the limitation to surface contribution in ARPES.

%------------------------------------------------------------------
\begin{figure*}
  \centering
  \includegraphics[width=1.0\textwidth]{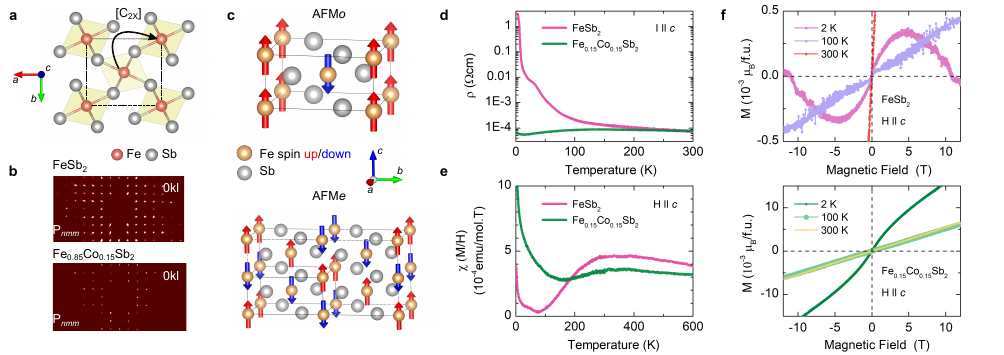}
  \small
  \caption{\textbf{Characterization of FeSb$_2$ and Fe$_{0.85}$Co$_{0.15}$Sb$_2$.}
    \textbf{a}~Crystal structure of FeSb$_2$ in the orthorhombic
    $Pnnm$ unit cell. \textbf{b}~Single-crystal X-ray diffraction patterns for undoped FeSb$_2$ and Fe$_{0.85}$Co$_{0.15}$Sb$_2$, confirming that
    moderate Co substitution preserves the $Pnnm$ structure. \textbf{c}~Altermagnetic (AFM$\textit{o}$) spin configuration and conventional antiferromagnetic (AFM$\textit{e}$) order resulting in a quadrupled unit cell. \textbf{d}~Temperature dependence of the electrical resistivity $\rho(T)$ and \textbf{e}~magnetic susceptibility $\chi(T)$ of undoped FeSb$_2$ and Fe$_{0.85}$Co$_{0.15}$Sb$_2$ along $c$-axis. \textbf{f}~Field-dependent magnetization $M(H)$ at 2~K, 100~K and 300~K for undoped FeSb$_2$ and Fe$_{0.85}$Co$_{0.15}$Sb$_2$ for field along the $c$-axis.}
  \label{fig1crystalstructure}
\end{figure*}
%------------------------------------------------------------------

Among candidate altermagnets, FeSb$_2$ is particularly compelling because its crystal symmetry promotes $d$-wave altermagnetic order, while first principle studies predict a metallic ground state, direction-dependent transport, and robustness under hydrostatic pressure~\cite{phillips2025electronic,bhandari2025effect,vsmejkal2022emerging,dou2025anisotropic, mazin2021prediction}. In its undoped form, however, FeSb$_2$ is a correlated narrow-gap semiconductor with an activated transport gap of about 30$~\mathrm{meV}$ and an optical gap of approximately 120$~\mathrm{meV}$\cite{petrovic2003anisotropy,perucchi2006optical,herzog2010strong,homes2018unusual}; it also exhibits a colossal thermoelectric Seebeck effect that cannot be captured by a simple band picture \cite{bentien2007colossal,sun2009huge}. 

Experimentally, no long-range magnetic order is observed in FeSb$_2$ \cite{zaliznyak2011absence,li2024spectroscopic,diakhate2011thermodynamic}. A consistent theoretical interpretation is provided by Takahashi’s spin-fluctuation framework, in which strong itinerant spin fluctuations suppress static magnetic ordering \cite{takahashi1997spin,koyama2010magnetization,el2024evolution}. These trends in the magnetic ground state were systematically examined by Kuhn \textit{et al.}~\cite{kuhn2023electronic}, who confirmed that FeSb$_2$ remains nonmagnetic with pronounced spin fluctuations, in contrast to isostructural CrSb$_2$ which orders antiferromagnetically. In fact, Mazin \textit{et al.} \cite{mazin2021prediction} showed that FeSb$_2$ sits near a magnetic instability: its lowest-energy states include both a conventional antiferromagnetic order~(AFM$\it{e}$)~and an altermagnetic~(AFM$\it{o}$)~order, as illustrated in Fig.~\ref{fig1crystalstructure}c.

The near-degeneracy of competing spin configurations in FeSb$_2$ reinforces the picture of a nonmagnetic ground state governed by strong spin fluctuations. At the same time, it suggests a possible tuning route: substituting Fe with Co introduces an additional $d$ electron that lowers the energy of the AFM$\it{o}$ configuration and may stabilize altermagnetic order \cite{mazin2021prediction}.

This scenario inspired our research on Co-doped FeSb$_2$; we present optical evidence that moderately doped FeSb$_2$ hosts a metallic altermagnetic state persisting up to room temperature. This metallicity is accompanied by microscopic lattice modifications that enhance electron-phonon coupling, while leaving the lattice vibrational modes largely unaffected by the altermagnetic order. These results demonstrate that carrier doping is a route to stabilize metallic altermagnetism in correlated materials with competing magnetic interactions.

%------------------------------------------------------------------
\section*{Results and Discussion}
%------------------------------------------------------------------
We first present the characterization of the parent compound FeSb$_2$ and the doped Fe$_{1-x}$Co$_{x}$Sb$_2$ systems by transport and magnetization measurements, before focusing on the electrodynamic properties. Particular emphasis is placed on the evolution of the interband transitions and vibrational features upon cobalt substitution and their explanation by altermagnetic spin ordering.

\subsection*{Doping semiconducting FeSb$_2$ towards a metallic state}
%------------------------------------------------------------------

FeSb$_2$ crystallizes in an orthorhombic structure (space group $Pnnm$), where Fe atoms are octahedrally coordinated by Sb [Fig.~\ref{fig1crystalstructure}a]. This structure hosts two Fe spin sublattices related by the combination of two-fold rotation and translation operation $[C_{2} \parallel C_{2x}t]$~\cite{holseth1968compounds,mazin2021prediction,dou2025anisotropic}, the defining symmetry of altermagnetism, making FeSb$_2$ a natural altermagnetic candidate. Moderate Co substitution ($\leq 50\%$ \cite{hu2006anisotropy}) preserves the crystal structure, as confirmed by single-crystal X-ray diffraction [Fig.~\ref{fig1crystalstructure}b].

\begin{figure}
  \centering
  \includegraphics[width=0.45\textwidth]{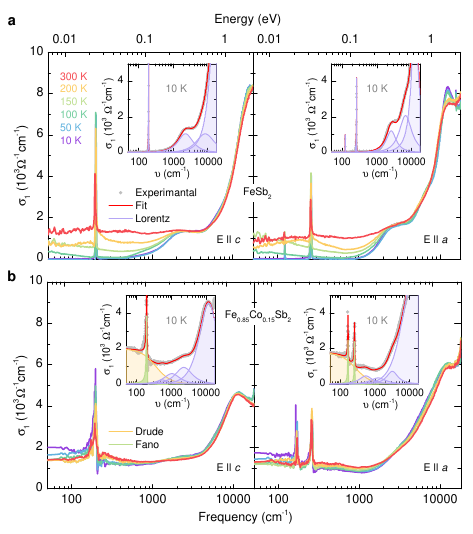}
  \small
  \caption{\textbf{Optical conductivity of FeSb$_2$ and Fe$_{0.85}$Co$_{0.15}$Sb$_2$.} \textbf{a}~Real part of the frequency-dependent optical conductivity $\sigma_1(\omega)$ of undoped FeSb$_2$ and \textbf{b}~Fe$_{0.85}$Co$_{0.15}$Sb$_2$ for light polarized along the $c$- and $a$-axes at selected temperatures. Insets: Drude-Lorentz decomposition at 10~K revealing the Drude response and low-energy interband transitions emerging upon Co doping.}
  \label{optcond}
\end{figure}

The electrical resistivity of FeSb$_2$ [Fig.~\ref{fig1crystalstructure}d] increases monotonically upon cooling, consistent with thermally activated transport across a narrow gap~\cite{petrovic2003anisotropy,bentien2007colossal}; this behavior resembles the correlated semiconductors FeSi \cite{schlesinger1993unconventional,Degiorgi1994FeSi} and SmB$_6$ \cite{cooley1995sm,gorshunov1999SmB6}, rather than a conventional band insulator. The optical conductivity [Fig.~\ref{optcond}a] corroborates this picture: the spectral weight, ${\rm SW} = \int \sigma _{1}(\omega){\rm d}\omega$, is thermally redistributed from low to high frequencies upon cooling (most prominently along the $a$- and $b$- axes \cite{homes2018unusual}), and an optical gap 2$\Delta\approx$ 110~meV is resolved along the $c$- and $a$-axes. The magnetic susceptibility, $\chi(T)$ [Fig.~\ref{fig1crystalstructure}e]~follows a thermally activated form with no anomalies associated with long-range magnetic order as established by inelastic neutron scattering \cite{zaliznyak2011absence}, confirming a non-magnetic ground state whose temperature evolution is governed by itinerant spin fluctuations \cite{koyama2010magnetization,li2024spectroscopic}.

Upon moderate Co substitution (here, "moderate" refers to $\sim$10--15~\% Co concentration and we focus on the 15~\%), FeSb$_2$ transitions into a metallic ground state. As the temperature is reduced, the resistivity of Fe$_{0.85}$Co$_{0.15}$Sb$_2$ [Fig.~\ref{fig1crystalstructure}d] initially tracks the thermally activated behaviour of the pristine compound, then drops sharply below $\sim$100~K, marking the crossover into a coherent metallic transport, consistent with previous reports \cite{hu2006anisotropy,Hu2007weakferromagnetism}. Concomitantly, the magnetic susceptibility undergoes a much shallower temperature variation [Fig. \ref{fig1crystalstructure}e] compared to the undoped compound: the strong temperature-dependent $\chi(T)$ that characterises the itinerant spin fluctuations of FeSb$_2$ is substantially quenched upon Co doping \cite{mazin2021prediction,koyama2010magnetization}.

%------------------------------------------------------------------
%------------------------------------------------------------------

Across the measured temperature range (2 to 1000~K), no sharp anomaly is resolved in $\chi(T)$. Apart from structural decomposition near 700~K~\cite{rosenqvist1953magnetic,holseth1968compounds}, only a broad feature (350~K) near saturation is observed
(similar to FeSb$_2$) whose origin remains open, see Fig. S11 \cite{SM}. The absence of a sharp anomaly in the uniform bulk susceptibility $\chi(T)$, however, does not preclude symmetry-driven long-range altermagnetic order \cite{vsmejkal2022emerging}, or conventional antiferromagnetism, like in CrSb$_2$, whose N\'{e}el temperature is masked by a strong paramagnetic background \cite{sales2012transport,barthem2013revealing}. At low temperatures, the upturn in $\chi(T)$ for both samples is well described by a Curie-Weiss contribution attributable to dilute Fe impurities, see Fig. S7 \cite{SM} and \cite{pokharel2013magnetic}, and therefore does not indicate an intrinsic magnetic instability of the Co-doped system.

The field-dependent magnetization $M(H)$ [Fig.~\ref{fig1crystalstructure}f] reflects the dominant response of itinerant carriers to the applied field. Down to 100~K, both FeSb$_2$ and Fe$_{0.85}$Co$_{0.15}$Sb$_2$ exhibit a strictly linear increase: $M \propto H$, consistent with Pauli paramagnetism. At the lowest temperatures, undoped FeSb$_2$ develops sublinear behavior $M(H)$ reflecting the onset of a diamagnetic background as itinerant Fe~3$d$ states are thermally depopulated \cite{li2024spectroscopic}, while Fe$_{0.85}$Co$_{0.15}$Sb$_2$ remains linear down to 2~K, in accord with the metallic ground state established upon Co doping. No hysteresis is observed at any temperature. The spontaneous moment, extracted by linear extrapolation of high-field $M(H)$ to zero, remains below  $3 \times 10^{-3}~\mu_B$/f.u. at $T=2$~K; hence ferromagnetic ordering can be excluded. The response is anisotropic, with a larger non-saturating magnetization along $H \parallel c$ than $H \parallel a$, suggesting contributions from spin degrees of freedom beyond the paramagnetic response of the itinerant carriers. A detailed analysis is given in Figs. S8 and S9~\cite{SM}.

%------------------------------------------------------------------
\subsection*{Carrier dynamics}
%------------------------------------------------------------------

While signatures of compensated magnetic ordering may be screened in bulk magnetization measurements, the electronic band structure is directly affected by the underlying spin configuration, leading to unique fingerprints in optical transitions \cite{ferber2010analysis,wenzel2025fermi,kopf2020influence,fang2013structural}. The optical conductivity of Fe$_{0.85}$Co$_{0.15}$Sb$_2$ [Fig.~\ref{optcond}b] differs markedly from the parent compound and remains anisotropic between $a$- and $c$-axes. A decomposition of $\sigma_1(\omega)$ at 10~K (inset of Fig.~\ref{optcond}b) reveals a Drude response together with several low-energy transitions centered around 0.1~eV; both are absent in the undoped FeSb$_2$. The emergence of the Drude component signals the closing of the narrow semiconducting gap, consistent with the metallic transport as Co doping increases. Upon lowering the temperature, spectral weight is weakly redistributed toward higher energies, in contrast to the pronounced transfer seen in undoped FeSb$_2$. It is crucial to note that no abrupt redistribution is observed at any temperature, establishing an electronically stable ground state below room temperature. Together, these observations set the stage for identifying the origin of the low-energy interband transitions emerging under Co-doping.
%------------------------------------------------------------------
\begin{figure*}
\centering
\begin{minipage}{0.36\textwidth}
\small
\caption{\textbf{Electronic band structure and optical conductivity of FeSb$_2$.}
\textbf{a}~Electronic band structure of nonmagnetic (NM) FeSb$_2$ calculated with PBE, mBJ, and PBE$+U$ exchange-correlation functionals. \textbf{b}~Electronic band structure of FeSb$_2$ in AFM$\textit{o}$ magnetic configuration, calculated using PBE exchange-correlation function (without SOC). \textbf{c}~Density of states (DOS) of nonmagnetic FeSb$_2$ for PBE, mBJ, and PBE$+U$ functionals, together with the DOS of the altermagnetic AFM$\textit{o}$ configuration (black dashed line). \textbf{d}~Calculated optical conductivity $\sigma_{zz}$ for the non-magnetic (NM) configuration for the cases shown in \textbf{a} and \textbf{b}. \textbf{e}~Comparison of experimental interband optical conductivity with DFT calculations for the non-magnetic structure along the $c$-axis and $a$-axis. Note that a Gaussian broadening of 0.05~eV was applied to DFT conductivity. \textbf{f}~Experimental conductivity $\sigma_1(\omega)$ along the $c$-axis for undoped FeSb$_2$ ($x = 0$) and Fe$_{1-x}$Co$_x$Sb$_2$ with $x = 0.05$, $0.12$, and $0.15$.}
\label{fesb2optdft}
\end{minipage}
\hfill
\begin{minipage}{0.60\textwidth}
\includegraphics[width=\textwidth]{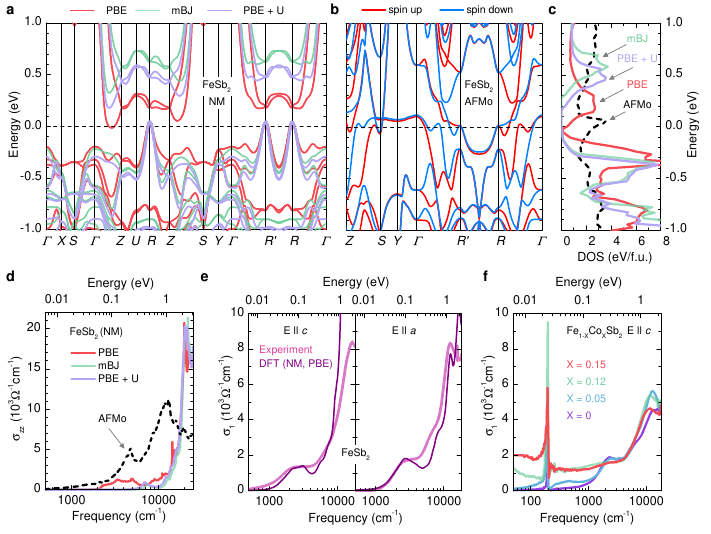}
\end{minipage}
\end{figure*}
%------------------------------------------------------------------

A quantitative analysis of the Drude response provides further insight into the nature of the itinerant carriers. At $T = 10$~K, the $c$-axis response of Fe$_{0.85}$Co$_{0.15}$Sb$_2$ is characterized by a scattering rate $1/\tau \approx 336$~cm$^{-1}$ and a screened plasma frequency $\omega_p^{\mathrm{scr}} \approx 6320$~cm$^{-1}$; along the $a$-axis, the extracted parameters are $1/\tau \approx 274$~cm$^{-1}$ and $\omega_p^{\mathrm{scr}} \approx 5400$~cm$^{-1}$. The corresponding unscreened plasma frequencies, $\omega_p = \omega_p^{\mathrm{scr}}\sqrt{\varepsilon_{\infty}}$, are 2.23 and 1.95~eV along the $c$- and $a$-axes, respectively. This anisotropy in both scattering rate and carrier density ($\omega_p^2 \propto n/m^*$) indicates direction-dependent dynamics of the itinerant charge carriers.

Strikingly, even at $T = 300~\mathrm{K}$, the Drude scattering rate of Fe$_{0.85}$Co$_{0.15}$Sb$_2$ is already reduced to half the value found in the parent compound at the same temperature. Specifically, $1/\tau$ decreases from $\approx 905~\mathrm{cm}^{-1}$ (along the $c$-axis) and $1221~\mathrm{cm}^{-1}$ (along $a$) in the undoped compound to $559~\mathrm{cm}^{-1}$ and $336~\mathrm{cm}^{-1}$, respectively, upon Co doping. This reduction occurs despite scattering off Co impurities. Since Co substitution not only closes the gap, but also simultaneously reduces the carrier scattering, we conclude that doping suppresses the spin fluctuations, which dominate incoherent scattering in the parent compound \cite{koyama2010magnetization,mazin2021prediction}. A comprehensive presentation of the optical spectra is provided in Figs. S1--S4 of the Supplemental Material~\cite{SM}.

%------------------------------------------------------------------
\subsection*{Pseudogap in FeSb$_2$}
%------------------------------------------------------------------

Assuming the absence of long-range magnetic order, density functional 
theory (DFT) calculations using the Perdew-Burke-Ernzerhof (PBE) exchange-correlation potential~\cite{PBE1996} predict an incipient 
metallic state for undoped FeSb$_2$ [see red curve in Fig.~\ref{fesb2optdft}a], consistent with previous reports~\cite{kuhn2023electronic, homes2018unusual,tomczak2025thermopower}. In this case, the conduction band barely crosses the Fermi level along the $\Gamma \rightarrow Z$ direction, resulting in a pseudogap with a vanishingly small density of states (DOS) at $E_\mathrm{F}$, as evidenced by the DOS shown in Fig.~\ref{fesb2optdft}c. Using the more advanced, modified Becke-Johnson (mBJ) exchange-correlation potential~\cite{Tran2009}, opens a direct band gap of approximately 0.5~eV [see Fig.~\ref{fesb2optdft}a], recovering a fully gapped semiconducting state. A similar trend is observed when accounting for the correlated nature of the Fe~$3d$ electrons within the DFT$+U$ framework ($U_\mathrm{Fe} = 2$~eV). Although PBE typically underestimates semiconductor gaps~\cite{Tran2009}, in FeSb$_2$ mBJ and DFT$+U$ overestimate the band gap, whereas PBE captures the experimental spectral weight realistically as discussed below. 

%------------------------------------------------------------------

\begin{figure*}
  \centering
  \includegraphics[width=1.0\textwidth]{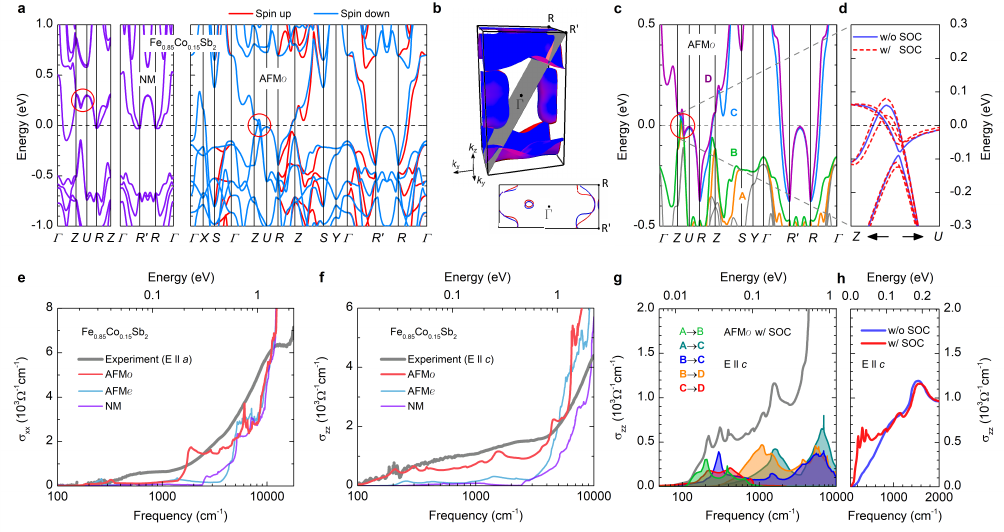}
  \small
\caption{\textbf{Electronic band structure and optical conductivity of Fe$_{0.85}$Co$_{0.15}$Sb$_2$.} \textbf{a}~Band structure of Fe$_{0.85}$Co$_{0.15}$Sb$_2$ in the non-magnetic case, along with the spin-split electronic band structure in the altermagnetic AFM$\textit{o}$ configuration, calculated without SOC. The spin-up and spin-down bands of AFM$\textit{o}$ are shown in red and blue, respectively. \textbf{b}~Fermi surface and cross-section illustrating the momentum-dependent spin splitting characteristic of altermagnetic order. \textbf{c}~Band structure with important bands color-labeled; their transitions are shown in the band-resolved conductivity in \textbf{g}. Magnified in \textbf{d}, are the bands along $Z$--$U$, with and without SOC. \textbf{e}~Experimental  conductivity $\sigma_1(\omega)$ of
Fe$_{0.85}$Co$_{0.15}$Sb$_2$ along the $a$-axis compared with DFT calculations for altermagnetic (AFM$\textit{o}$), conventional antiferromagnetic (AFM$\textit{e}$) and non-magnetic (NM) configurations. \textbf{f}~Same as \textbf{e} for the $c$-axis. \textbf{g} Band-resolved optical conductivity with SOC, highlighting the low-energy optical transitions. \textbf{h} The low-energy optical conductivity of AFM$\textit{o}$ with and without SOC.}
\label{Cofesb2opt}
\end{figure*}
%------------------------------------------------------------------

In the calculated optical conductivity presented in Fig.~\ref{fesb2optdft}d, while both mBJ and DFT$+U$ approaches lead to a fully gapped ground state, they fail to reproduce the absorption in the range 1000--8000~cm$^{-1}$ observed in the experimental optical conductivity ($T = 10$~K) [see Fig.~\ref{fesb2optdft}e]; the onset of optical transitions in these gapped solutions is pushed beyond $\sim$7000~cm$^{-1}$, far exceeding the experimental zero-spectral weight region. The simple PBE approach, by contrast, accurately reproduces the experimentally observed interband optical transitions along the $c$-axis and $a$-axis, with a moderate rescaling of the energy axis (by a factor of $1/1.8$ and $1/1.2$ ), consistent with an anisotropic band renormalization in the semiconducting structure of FeSb$_2$~\cite{homes2018unusual}. The simultaneous success of PBE and failure of the static-potential mBJ and on-site-Coulomb DFT$+U$ corrections indicate that the gap in FeSb$_2$ is not well captured by the conventional static mean-field approach, pointing instead to the role of dynamical correlations and itinerant spin fluctuations. This scenario corroborates FeSb$_2$ as a correlated semiconductor with a narrow gap and strong spin fluctuations~\cite{tomczak2025thermopower,wenzel2025fermi,Mazin1999}.

Including the altermagnetic AFM$\textit{o}$ order (without SOC) in undoped FeSb$_2$ yields spin-split bands with several crossings at $E_{\mathrm{F}}$ [Fig.~\ref{fesb2optdft}b] and a finite DOS at $E_{\mathrm{F}}$ [Fig.~\ref{fesb2optdft}c]. Despite this metallicity in DOS, the optical conductivity [Fig.~\ref{fesb2optdft}d] retains only weak low-energy spectral weight ($<2000$~cm$^{-1}$) and develops pronounced interband transitions relative to the non-magnetic case, as a consequence of multiple bands lying close to $E_{\mathrm{F}}$. The closer agreement of the non-magnetic optical response with experiment identifies the non magnetic configuration as the ground state of undoped FeSb$_2$, and suggests that a metallic electronic structure is a prerequisite for AFM$\textit{o}$ order to stabilize.

%------------------------------------------------------------------
\subsection*{Metallic state and altermagnetic order due to Co doping}
%------------------------------------------------------------------

Figure~\ref{fesb2optdft}f shows the optical conductivity of FeSb$_2$ measured at $T=10$~K for various Co concentrations. The most pronounced changes occur in the low-energy spectral range: new interband transitions near 0.1~eV emerge already at 5\% Co doping and grow for increasing substitution; the Drude response appears at 12\% Co and strengthens further at 15\%. This doping-dependent evolution demonstrates a systematic, carrier-tuned reconstruction of the low-energy electronic structure. To compare directly with the DFT-calculated optical conductivity, we focus on the interband contribution of Fe$_{0.85}$Co$_{0.15}$Sb$_2$, obtained by subtracting the Drude component from the experimental $\sigma_1(\omega)$.  

The experimental interband optical conductivity of Fe$_{0.85}$Co$_{0.15}$Sb$_2$ along the $a$- and $c$-axes is shown in Figs.~\ref{Cofesb2opt}e,f, together with the calculated optical conductivity for altermagnetic (AFM$\textit{o}$), conventional antiferromagnetic (AFM$\textit{e}$) and non-magnetic (NM) configurations. Notably, no energy rescaling was applied, indicating that the effect of itinerant spin fluctuations is suppressed upon Co doping. Comparing with the experimental optical conductivity, neither the NM nor the AFM$\textit{e}$ configuration reproduces the observed spectral weight; both yield similar spectra inconsistent with experiment. The AFM$\textit{o}$ configuration, by contrast, exclusively reproduces the experimental spectra along both axes, providing unambiguous optical evidence for the AFM$\textit{o}$ altermagnetic electronic structure in Fe$_{0.85}$Co$_{0.15}$Sb$_2$.  

The microscopic origin of the AFM$\textit{o}$-induced optical response is revealed by the electronic structure. In the non-magnetic configuration [Fig.~\ref{Cofesb2opt}a], Fe$_{0.85}$Co$_{0.15}$Sb$_2$ retains a band structure similar to undoped FeSb$_2$ [Fig.~\ref{fesb2optdft}a], with a rigid band shift toward $E_\mathrm{F}$ and minor band modifications. Upon including the AFM$\textit{o}$ order (without SOC), spin-split bands appear near $E_\mathrm{F}$ along $Z$--$S$, $\Gamma$--$R'$, and $R$--$\Gamma$ directions, with a splitting as large as $\sim$0.2~eV; band crossings emerge along $\Gamma$--$X$, $Z$--$U$ and $R$--$Z$ directions. The Fermi surface [Fig.~\ref{Cofesb2opt}b] consists of spin-split sheets separated by nodal planes, rather than nodal points as in the case of pure SOC, consistent with the symmetry requirements of altermagnetism~\cite{vsmejkal2022emerging}. Moreover, the momentum-space cross-section reveals alternating spin-up and spin-down Fermi surface sheets such that the net magnetization vanishes~\cite{mazin2021prediction}. 

%------------------------------------------------------------------
\begin{figure*}[!ht]
\centering
\begin{minipage}[t]{0.35\textwidth}
  \vspace{0pt}
  \small
  \caption{\textbf{Lattice dynamics of undoped and Co-doped FeSb$_2$.}
  \textbf{a}~Infrared optical conductivity $\sigma_1(\omega)$ at 10~K showing the infrared-active phonon modes of undoped FeSb$_2$ and 
  Fe$_{0.85}$Co$_{0.15}$Sb$_2$ along the $c$- and $a$-axes; Fano lineshapes emerging upon Co substitution are highlighted by the asymmetric fits. \textbf{b}~Raman intensity at 10~K showing the Raman-active modes of the same compounds along the $c$- and $a$-axes. \textbf{c}~Calculated phonon dispersion of nonmagnetic (NM) FeSb$_2$ and \textbf{d}~Fe$_{0.85}$Co$_{0.15}$Sb$_2$ in the AFM$\textit{o}$ configuration, revealing zone-boundary imaginary frequencies and frequency hardening relative to the undoped nonmagnetic FeSb$_2$.}
  \label{fig4phonons}
\end{minipage}
\hfill
\begin{minipage}[t]{0.62\textwidth}
  \vspace{0pt}
  \includegraphics[width=\textwidth]{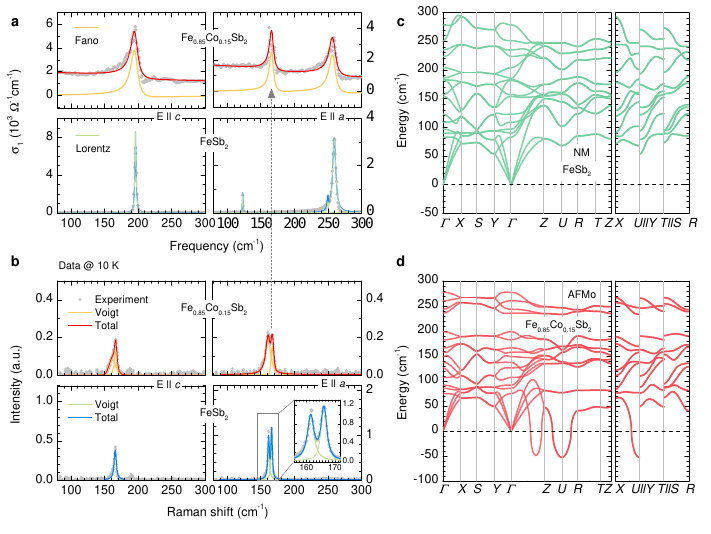}
\end{minipage}
\end{figure*}
%------------------------------------------------------------------

The origin of the low-energy spectral weight is identified through the band-resolved optical conductivity [Fig.~\ref{Cofesb2opt}g]: transitions below $\sim$1000~cm$^{-1}$ arise from the band crossings along the $Z$--$U$ and $R$--$Z$ directions [Fig.~\ref{Cofesb2opt}c], which become active exclusively through AFM$\textit{o}$ order. The corresponding bands in the $Z$--$U$ region lie $\sim$0.25~eV above $E_\mathrm{F}$ in the non-magnetic case [highlighted by red circles in Fig. \ref{Cofesb2opt}a]. Upon inclusion of SOC, these bands split by $\sim$5~meV along $Z$--$U$ [Fig.~\ref{Cofesb2opt}d], activating sharper low-energy transitions [Figs.~\ref{Cofesb2opt}g,h], while the dominant altermagnetic spin splitting of $\sim$0.2~eV remains unaffected, confirming its non-relativistic exchange origin. Since the optical selection rule $\Delta S = 0$ forbids direct transitions between spin-split bands~\cite{dressel2002electrodynamics,Bassani1975}, the low-energy optical activity originates from the metallic band crossings along $Z$--$U$ and $R$--$Z$; for instance, the C$\rightarrow$D transition centered around $\sim$30~meV [Fig.~\ref{Cofesb2opt}g]. 

%------------------------------------------------------------------
\subsection*{Lattice dynamics affected by Co substitution and magnetic ordering}
%------------------------------------------------------------------

Co substitution in FeSb$_2$ is also reflected in the lattice dynamics, providing complementary evidence for electronic reconstruction. Figs.~\ref{fig4phonons}a,b show the infrared optical conductivity and Raman intensity of the phonon modes at 10~K for undoped and 15\% Co-doped FeSb$_2$. For light polarized along the $c$- and $a$-axes, three infrared-active modes are observed in undoped FeSb$_2$: $B_{3u}$, $B_{1u}$, and $B_{2u}$. In addition, the relevant Raman-active modes A$_g$ and $B_{1g}$ are present. The phonon peaks were fitted using Lorentzian and Voigt profiles, yielding symmetric lineshapes for the undoped compound, consistent with the absence of significant electron-phonon coupling in the semiconducting parent \cite{homes2018unusual,Lazarević2012lattice}.

Upon moderate Co substitution, the infrared-active phonons acquire pronounced asymmetric Fano lineshapes and broaden substantially, signaling strongly enhanced electron-phonon coupling in the metallic phase~\cite{fano1961effects}. In Raman spectroscopy, the same electron-phonon coupling manifests as symmetric broadening of the Raman-active modes and mode-mixing, without producing a visible Fano asymmetry [Fig.~\ref{fig4phonons}b], as expected for the weaker electronic Raman background in this frequency range~\cite{lazarevic2009raman}. The temperature and doping dependence of these phonon modes shows continuous hardening, ruling out any structural transition and confirming that the system remains in a single homogeneous phase up to room temperature. Detailed phonon spectra are shown in Figs.~S16--S18 and Tables~\text{II, III} of the Supplemental Material~\cite{SM}.

%------------------------------------------------------------------
\begin{table*}[t]
\small
\caption{Experimental and calculated phonon frequencies of undoped FeSb$_2$ and Fe$_{0.85}$Co$_{0.15}$Sb$_2$. Experimental frequencies are listed for infrared-active (IR) and Raman-active (R) modes measured at 10~K. Calculated frequencies are obtained for the nonmagnetic (NM) undoped compound and for 15\% Co-doped FeSb$_2$ in the nonmagnetic, altermagnetic (AFM$\textit{o}$), and conventional antiferromagnetic (AFM$\textit{e}$) configurations. All frequencies are given in cm$^{-1}$.R (IR) represents the Raman-active mode in undoped FeSb$_2$ that becomes additionally infrared-active upon Co substitution.}
\label{tab:phonons}
\centering
\setlength{\tabcolsep}{7pt}
\begin{tabular}{lccccccccc}
\toprule
 & & & \multicolumn{2}{c}{Experiment} & &
   \multicolumn{4}{c}{Calculation} \\
\cmidrule(lr){4-5} \cmidrule(lr){7-10}
Mode & Polarization & Activity &
FeSb$_2$ & Fe$_{0.85}$Co$_{0.15}$Sb$_2$ & &
NM & 15\% Co & AFM$\textit{o}$ (with 15\% Co) & AFM$\textit{e}$ \\
\midrule

$B_{3u}$ & $E \parallel a$ & IR
          & 123.2 & - & & 123.22 & 125.48 & 126.68 & 123.05 \\
$A_{g}$ & $E \parallel a$ & R
          & 161.55 & 161.90 & & 156.0 & 157.974 & 148.934 & 157.30 \\
$B_{1g}$  & $E \parallel a$ & R (IR)
          & 166.14 & 167.32 (166.7) & & 178.15 & 179.18 & 166.379 & 174.82 \\
$B_{1u}$  & $E \parallel c$ & IR
          & 195.61 & 196.37 & & 186.46 & 188.528 & 182.657 & 188.16 \\
$B_{2u}$  & $E \parallel a$ & IR
          & 249.98 & 257.8 & & 259.61 & 276.54 & 277.722 & 272.218 \\
\bottomrule
\end{tabular}
\begin{flushleft}
\end{flushleft}
\end{table*}
%------------------------------------------------------------------

Interestingly, the $B_{1g}$ mode that is Raman-active in undoped FeSb$_2$ and previously identified as carrying the strongest electron-phonon coupling among all phonon modes~\cite{Lazarevi2010evidence} becomes infrared-active upon Co substitution [Fig.~\ref{fig4phonons}a,b, indicated by an arrow]. Since the average crystal structure remains $Pnnm$, this mode activation points towards a local breaking of inversion symmetry 
at Co substitution sites~\cite{barker1975optical}, without disrupting the global altermagnetic spin symmetry. We further note that the B$_{3u}$ mode is screened by the Drude contribution in the metallic phase and is therefore not resolved in the infrared spectra of the doped compound. The relatively sharp phonon lineshapes and continuous evolution with temperature together support a microscopically uniform Co distribution that stabilizes the altermagnetic phase robustly against local lattice perturbations. We note, however, that optically visible surface inclusions in the doped crystals produce distinct Raman signatures while leaving the infrared phonon response unaffected. This mirrors the caution raised for $\alpha$-MnTe, where surface inclusions produced extrinsic Raman modes while leaving 
bulk infrared measurements unaffected~\cite{uykur2026revisiting, 
belashchenko2026mystery}: see Fig. S20~\cite{SM}.

To examine the influence of magnetic ordering on lattice dynamics~\cite{steward2023dynamic, schilberth2025optical}, we calculated phonon frequencies for NM, AFM$\it{e}$ and AFM$\it{o}$ at the experimental Co concentration, with selected modes summarized in Table~\ref{tab:phonons} (details in Fig. S19 and Table. \text{IV} of the SM\cite{SM}). The calculated phonon frequencies of undoped FeSb$_2$ are in reasonable agreement with the experiment. Co substitution generally hardens the phonon frequencies through bond stiffening, while the calculated frequency shifts between magnetic configurations suggest a weak magnetoelastic coupling whose magnitude remains indistinguishable in experiment. For instance, the $B_{1u}$ mode shifts from 186.46~cm$^{-1}$ in the undoped NM configuration to 182.65~cm$^{-1}$ upon 15\% Co substitution, with a further minute shift to 182.79~cm$^{-1}$ in the AFM$\textit{o}$ configuration; the AFM$\textit{e}$ configuration yields 188.13~cm$^{-1}$. This suggests that the zone-center phonon frequencies are only weakly sensitive to the magnetic configuration, and that the dominant effect on lattice dynamics is electronic rather than magnetic in origin, with the frequency shifts and mode mixing primarily attributable to bond modifications associated with Co substitution.

Furthermore, the AFM$\textit{o}$ phonon dispersion of Fe$_{0.85}$Co$_{0.15}$Sb$_2$ [Fig.~\ref{fig4phonons}d] reveals imaginary phonon frequencies along $\Gamma$--$Z$ and around the $Z$--$U$--$R$ region of the Brillouin zone, absent in undoped FeSb$_2$ [Fig.~\ref{fig4phonons}c]. This is the same region where the electronic band structure exhibits band crossing [Fig.~\ref{Cofesb2opt}a], enforced by the AFM$\textit{o}$ order. Since the crystal structure remains $Pnnm$, these imaginary frequencies do not indicate an intrinsic structural instability. Rather, the momentum-space coincidence with the band crossings points to an enhanced electron-phonon coupling at these wavevectors~\cite{xue2019electron}. These zone-boundary features are consistent with the same enhanced electron-phonon coupling that manifests at the zone center as the Fano lineshapes and mode mixing observed experimentally, establishing a coherent picture of electron-phonon coupling across the Brillouin zone. 

%------------------------------------------------------------------
\section*{Conclusion}
%------------------------------------------------------------------

We have demonstrated that moderate Co substitution ($\sim$15\%) drives the correlated narrow-gap semiconductor FeSb$_2$ into a metallic altermagnetic state that persists up to room temperature. The carrier-induced closure of the correlation-driven narrow-gap is the critical precondition for this transition: it simultaneously suppresses the strong spin fluctuations that prevented long-range magnetic order in the parent compound and enables distinguishable optical activity in the altermagnetic band structure. The resulting low-energy interband transitions near 0.1~eV, absent in undoped FeSb$_2$, are reproduced exclusively by the AFM$\textit{o}$ configuration in DFT without the need for band renormalization, suggesting the suppression of strong spin fluctuations observed in the undoped compound. This match constitutes unambiguous bulk optical evidence for metallic altermagnetism in Fe$_{0.85}$Co$_{0.15}$Sb$_2$, ruling out conventional antiferromagnetic and non-magnetic electronic structures. Crucially, the dominant altermagnetic spin splitting of $\sim$0.2~eV is of non-relativistic exchange origin, while spin-orbit coupling enters only as a perturbative correction of $\sim$5~meV at the band crossing along the $Z$--$U$ direction. 

The lattice dynamics provide independent and complementary evidence for the electronic reconstruction. Co substitution activates the Raman-active B$_{2g}$ mode in the infrared spectrum, along with pronounced Fano lineshapes, providing fingerprints of strongly enhanced electron-phonon coupling. The calculated phonon dispersion of Fe$_{0.85}$Co$_{0.15}$Sb$_2$ reveals zone-boundary imaginary frequencies in the $\Gamma$--$Z$ and $Z$--$U$--$R$ regions, precisely the same regions where band crossing appears in the AFM$\textit{o}$ electronic band structure. This metallicity, induced by Co substitution, plays a crucial role in stabilizing the altermagnetic order in moderately doped FeSb$_2$.   

More broadly, our results identify carrier-tuned FeSb$_2$ as a model platform for metallic $d$-wave altermagnetism, where the competing magnetic configurations and the strength of electron-phonon coupling can be systematically controlled through Co concentration. The realization of a metallic altermagnetic bulk state in this system suggests carrier tuning as an effective strategy to stabilize metallic altermagnetism in correlated semiconductors with competing magnetic interactions. Together, the bulk-sensitive optical conductivity and DFT calculations establish a powerful route to identify and confirm altermagnetic order in quantum materials, complementing surface-sensitive probes such as ARPES and avoiding extrinsic contributions inherent to transport measurements. The predicted large anomalous Hall conductivity and magneto-optical Kerr effect~\cite{mazin2021prediction} in the altermagnetic phase, as well as the non-relativistic spin-splitter effect accessible via spin Hall measurements, provide immediate experimental targets that can now be pursued in this metallic system.

%------------------------------------------------------------------
\section*{Methods}
%------------------------------------------------------------------
\subsection*{Crystal growth}
%------------------------------------------------------------------
Single crystals of Fe$_{1-x}$Co$_x$Sb$_2$ with mm-sized mirror-like facets were grown from Sb-rich self-flux. Fe pieces (Puratronic, 99.995\%), Sb pieces (Alfa Aesar, 99.9999\%) and Co powder (Thermo Scientific, 99.998\%) were weighed into Canfield-type alumina crucibles and sealed in quartz ampoules, which were heated to maximum temperatures of 1000--1200~$^\circ$C before being slowly cooled at a rate of 0.5--0.7~$^\circ$C/h in the growth window from 760~$^\circ$C to 650~$^\circ$C and finally decanted with the help of a centrifuge. The pure FeSb$_2$ was grown out of a melt with composition Fe:Sb~$= 1:15.7$ and the Fe$_{0.85}$Co$_{0.15}$Sb$_2$ crystal was grown out of a melt with composition Fe:Co:Sb~$= (1-x):x:11.5$,$x=0.15$. Energy-dispersive X-ray spectroscopy (EDS) shows a Co content of $x_\mathrm{EDS} = 0.14(1)$, close to the nominal value. Other crystals of Fe$_{1-x}$Co$_x$Sb$_2$ were also grown out of a melt with Fe:Co:Sb~$= (1-x):x:11.5$ composition with $x = 0, 0.054, 0.115$. EDS shows Co contents of $x_\mathrm{EDS} = 0.025$ and $x_\mathrm{EDS} = 0.10$ for the latter two batches. 
%------------------------------------------------------------------
\subsection*{Transport and magnetic measurements}
%------------------------------------------------------------------
The temperature-dependent electrical resistivity was measured using a Physical Property Measurement System (PPMS, Quantum Design) from 300~K down to 2~K on the same crystals used for the optical measurements. All resistivity data are presented in Fig. S10~\cite{SM}. The temperature-dependent magnetic susceptibility was measured using a Vibrating Sample Magnetometer (VSM) under a fixed external magnetic field of 0.1~T. Field-dependent magnetization measurements were performed for fields of $\pm$12~T. High-temperature susceptibility was measured up to 1000~K using the oven insert option of the VSM.

%------------------------------------------------------------------
\subsection*{Infrared spectroscopy}
%------------------------------------------------------------------
The infrared reflectivity of polished crystals with an approximate reflecting-plane size of atleast $2 \times 1~\mathrm{mm}^2$ was measured in the frequency range 40--18{,}000~cm$^{-1}$ (5~meV to 2.2~eV) as a function of temperature from $T = 300$ to 10~K. A Bruker IFS 113v Fourier-transform infrared spectrometer was used for the far-infrared range ($\omega/2\pi c < 650$~cm$^{-1}$), and a Bruker Vertex 80v spectrometer attached to a Hyperion IR microscope was used for the mid- and near-infrared ranges ($\omega/2\pi c > 650$~cm$^{-1}$). The gold overcoating technique was applied to obtain the absolute reflectivity in the far-infrared range, while a gold mirror was used as a reference in the mid- and near-infrared ranges. The complex optical conductivity $\sigma(\omega) = \sigma_1(\omega) + \mathrm{i}\sigma_2(\omega)$ was obtained via Kramers-Kronig analysis. The low-frequency reflectivity below 50~cm$^{-1}$ was extrapolated with a Drude function for the metallic states and a constant for the insulating ones, while an X-ray atomic scattering function was used at high frequencies. We note here that the optical conductivity of FeSb$_2$ shown in Fig. \ref{optcond}a and Fig. \ref{fesb2optdft}e are from an annealed sample, which has sharper interband and phonon features than the non-annealed \ref{fesb2optdft}f. Further details are given in the sections S1 and S2 and Figs. S1--S6 of the Supplemental Material \cite{SM}.

%------------------------------------------------------------------
\subsection*{Raman spectroscopy}
%------------------------------------------------------------------
Polarized Raman spectra were acquired using a HORIBA Jobin-Yvon LabRAM HR800 spectrometer and a He-Ne laser with a wavelength of 632.8 nm. An open-cycle KONTI micro cryostat from CryoVac was used to access low temperatures.

%------------------------------------------------------------------
\subsection*{XRD and EDX characterization}
%------------------------------------------------------------------
We have characterized the exact crystals from optical spectroscopy 
measurements via a combination of single-crystal XRD and electron 
microscopy combined with Energy-Dispersive X-ray (EDX). The single-crystal XRD data was collected at room temperature with a Rigaku XtaLab Mini II Single Crystal X-ray Diffractometer using Mo-K$_{\alpha}$ radiation ($\lambda$~=~0.71073~\AA), graphite monochromator, operated 50~kV and 12~mA 600~W. The BSE imaging and EDX measurements were done with a JSM-IT210LA, with an accelaration voltage of 5 kV for imaging and 20 kV for EDX analysis. On every crystal a minimum of 6 points was taken to acquire statistical evidence. XRD and EDX spectra are shown in Figs. S12--S15 of the Supplemental Material \cite{SM}.
%------------------------------------------------------------------
\subsection*{Computational Details}
%------------------------------------------------------------------

Density functional theory (DFT) calculations were performed using the \texttt{Wien2k} code~\cite{wien2k, Blaha2020} and the Perdew-Burke-Ernzerhof (PBE) exchange-correlation functional~\cite{PBE1996}, unless stated otherwise. The DFT$+U$ was calculated using the FLL (fully localized limit) correction and $U_\mathrm{Fe} = 2$~eV. Self-consistent calculations were converged using a $10 \times 10 \times 20$ $k$-point mesh for the nonmagnetic case and the AFM$\textit{o}$ order, while calculations including the AFM$\textit{e}$ order employed a $17 \times 7 \times 15$ $k$-point mesh. Experimentally determined lattice parameters at 300~K, provided in Table \text{I} in the Supplemental Material~\cite{SM}, were used for all calculations. Co-doping was modelled using the virtual crystal approximation (VCA). Spin-orbit coupling (SOC) was included in all calculations unless stated otherwise. The optical conductivities were calculated using the \texttt{OPTIC} module~\cite{Draxl2006} on denser $k$-point meshes: a $39 \times 34 \times 71$ mesh was used for the nonmagnetic and AFM$\textit{o}$ cases, while a $36 \times 16 \times 33$ mesh was employed for the AFM$\textit{e}$ case.

Phonon dispersions were calculated based on density functional perturbation theory (DFPT), as implemented in the \texttt{PHONOPY} code, with a plane-wave cutoff energy of 520~eV and a $k$-point mesh of $1 \times 1 \times 1$. Supercells of size $3 \times 3 \times 3$ were adopted for the NM and AFM$\textit{o}$ configurations, whereas a $2 \times 2 \times 2$ supercell was used for the AFM$\textit{e}$ structure.
%------------------------------------------------------------------
\section*{Acknowledgements}
%------------------------------------------------------------------

We acknowledge insightful discussions with Igor Mazin (George Mason University, Virginia). We thank Achyut Tiwari (Universit\"at Stuttgart) for continuous discussions, A. V. Pronin (Universit\"at Stuttgart) for supervision, and Run Yang (Southeast University, Nanjing) for triggering the project. We thank M.~Minola and B. Keimer (Max Planck Institute for Solid State Research, Stuttgart) for providing access to Raman facility. We are grateful to Sarah Parks and Gabriele Untereiner for technical support. Crystal growth in Bochum was supported by the DFG under CRC/TRR~288 (Project A02).

%------------------------------------------------------------------
\section*{Author Contributions}
%------------------------------------------------------------------

R.M.R.\ and M.Po.\ conducted the optical measurements. R.M.R.\ wrote the manuscript with input from all authors. M.W.\ performed the DFT calculations and contributed to the analysis and discussion. J.K., M.vL., A.K., and A.B.\ synthesized the crystals. V.S.\ and R.M.R.\ carried out the Raman measurements. P.P.\ characterized the crystals with XRD\@. C.P.\ and M.Pi.\ contributed to transport and magnetic measurements. Y.X.\ and T.Z.\ performed the phonon calculations. M.D.\ supervised the whole project. All authors discussed the results and contributed to the manuscript.

%------------------------------------------------------------------
\section*{Data Availability}
%------------------------------------------------------------------
The data and files used for this study are available from the corresponding authors upon request.

%------------------------------------------------------------------
\section*{Competing Interests}
%------------------------------------------------------------------

The authors declare no competing interests.

%------------------------------------------------------------------
\bibliography{FeSb2}
%------------------------------------------------------------------

\end{document}